\documentclass [11pt]{article}

\usepackage{doublespace}
\usepackage{epsfig}
\usepackage{curves}
\usepackage{amssymb}
\usepackage{amsthm}
\usepackage{verbatim}
\usepackage{delarray}
\usepackage{graphics}
\usepackage{epic}

\newcommand{\ket}[1]{{|#1\rangle}}
\newcommand{\bra}[1]{{\langle#1|}}
\newcommand{\braket}[2]{{\langle#1|#2\rangle}}

\newcommand{\bl}{\left(}
\newcommand{\br}{\right)}

\newcommand{\tr}{\mbox{Tr}} 
\newcommand{\mvec}{\mbox{vec}}

\newcommand{\inner}[2]{\langle{#1},{#2}\rangle}

\newcommand{\hpi}{\widehat{\Pi}}
\newcommand{\hx}{\widehat{X}}

\newcommand{\HH}{{\mathcal{H}}}
\newcommand{\N}{{\mathcal{N}}}
\newcommand{\B}{{\mathcal{B}}}
\newcommand{\R}{{\mathcal{R}}}
\newcommand{\LL}{{\mathcal{L}}}

\newcommand{\ie}{{\em i.e., }}
\newcommand{\eg}{{\em e.g., }}

\newcommand{\etal}{\emph{et al.\ }}

\newtheorem{theorem}{Theorem}

\title{\singlespace Designing Optimal Quantum Detectors \\ Via Semidefinite
Programming} \author{Yonina C. Eldar\footnote{Research Laboratory of
Electronics, Massachusetts Institute of Technology, Room 36-615,
Cambridge, MA 02139. E-mail: yonina@alum.mit.edu.}, Alexandre
Megretski\footnote{Laboratory for Information and Decision Systems,
Massachusetts Institute of Technology, Cambridge, MA 02139. E-mail:
ameg@mit.edu.},  and George C. Verghese\footnote{
Laboratory for Electromagnetic and Electronic Systems,
Massachusetts Institute of Technology,
Cambridge, MA 02139. E-mail: verghese@mit.edu.}
}

\date{\today}
 
\begin{document} 

\maketitle

\renewcommand{\thefootnote}{}
\footnotetext{
This work is supported in part by BAE Systems Cooperative Agreement
RP6891 under Army Research Laboratory Grant DAAD19-01-2-0008, by the
Army Research Laboratory Collaborative Technology Alliance through BAE
Systems Subcontract RK78554, and by Texas Instruments through the TI
Leadership University Consortium.}
\renewcommand{\thefootnote}{\arabic{footnote}}

%%%%%%%%%%%%%%%%%%%%%%%%%%
\begin{abstract}
%%%%%%%%%%%%%%%%%%%%%%%%%%

\singlespace 
We consider the problem of designing an optimal quantum detector to
minimize the probability of a detection error when distinguishing
between a collection of quantum states, represented by a set of
density operators.  We show that the design of the optimal detector
can be formulated as a semidefinite programming problem. Based on this
formulation, we derive a set of necessary and sufficient conditions
for an optimal quantum measurement.  We then show that the optimal
measurement can be found by solving a standard (convex) semidefinite
program followed by the solution of a set of linear equations or, at
worst, a standard linear programming problem. By exploiting the many
well-known algorithms for solving semidefinite programs, which are
guaranteed to converge to the global optimum, the optimal measurement
can be computed very efficiently in polynomial time.

Using the semidefinite programming formulation, we also show that the
rank of each optimal measurement operator is no larger than the rank
of the corresponding density operator. In particular, if the quantum
state ensemble is a pure-state ensemble consisting of (not necessarily
independent) rank-one density operators, then we show that the optimal
measurement is a pure-state measurement consisting of rank-one
measurement operators.

\end{abstract}

\vspace{.5\baselineskip}\noindent {\em Index Terms\/}---Quantum detection,
semidefinite programming, duality.

%%%%%%%%%%%%%%%%%%%%%%%%%%
\section{Introduction}
\label{sec:intro}
%%%%%%%%%%%%%%%%%%%%%%%%%%

In a quantum detection problem a transmitter conveys
classical information to 
a receiver using a quantum-mechanical channel. Each
message is represented by preparing the quantum channel in a 
quantum state represented by a density operator, drawn from a collection
of known states.
At the receiver, the information is detected by subjecting the channel
to a quantum measurement 
in order to determine the state prepared. 
If the quantum states are mutually orthogonal, 
then the state can be determined correctly with
probability one by performing 
an optimal orthogonal (von Neumann) 
measurement \cite{P95}.
However, if the given states are not orthogonal,
then no measurement will  distinguish perfectly between them. 
Our  problem is therefore to construct a
measurement that minimizes the
probability of a detection error.

We consider a quantum state ensemble consisting of $m$ 
density operators $\{\rho_i, 1
\le i \le m\}$ on an $n$-dimensional complex Hilbert space $\HH$, with 
prior probabilities $\{p_i>0, 1 \le i \le m\}$.
A density operator $\rho$ is a 
positive
semidefinite (PSD)
 Hermitian operator with $\tr(\rho)=1$; we write $\rho \geq 0$ to indicate $\rho$ is PSD.
A pure-state ensemble
is one in which each density operator $\rho_i$ is a rank-one projector
$\ket{\phi_i}\bra{\phi_i}$, where the vectors $\ket{\phi_i}$, though
evidently normalized to unit length, are
 not necessarily orthogonal.

For our {\em measurement} we  consider general
positive operator-valued measures \cite{H76,P90},
consisting of $m$ PSD
Hermitian operators $\{\Pi_i, 1 \le i \le m\}$ that form a resolution
of the identity on $\HH$.
A pure-state measurement is one in
which each measurement operator $\Pi_i$ is a rank-one operator\footnote{In this paper, when we say rank-one
operator we mean an operator that can be expressed in the form
$\Pi_i=\ket{\mu_i}\bra{\mu_i}$ for some  $\ket{\mu_i} \in
\HH$. Note, however, that $\ket{\mu_i}$ may be equal to $0$ in which
case the operator actually has rank zero.} 
$\ket{\mu_i}\bra{\mu_i}$, where the vectors $\ket{\mu_i}$ are not
necessarily orthogonal or normalized.  An orthogonal measurement (\ie a von
Neumann measurement) is one in which
the measurement operators $\Pi_i$ are mutually orthogonal projection operators.

Necessary and sufficient conditions for an optimum measurement
minimizing the probability of a detection error
have been derived \cite{H73,YKL75}.
However, except in some particular cases \cite{H76,CBH89,OBH96,BKMO97,EF01}, 
obtaining a closed-form analytical expression for the optimal
measurement directly from these conditions is 
a difficult and unsolved problem. Thus in practice, 
iterative procedures 
\cite{H82} or {\em ad-hoc} suboptimal measurements are used.
A detection measurement that has many desirable properties and has
been employed in 
many settings is the
least-squares measurement \cite{EF01}, also known as the 
square-root measurement \cite{HW94,H96}.

Holevo \cite{H73} derives the necessary and sufficient conditions by
considering infinitesimal 
transformations of the measurement operators $\Pi_i$ that preserve
their character as elements of a measurement. The drawback of this approach is
that it does not readily lend itself to efficient computational
algorithms. Yuen \etal \cite{YKL75} use the principle of duality  in
vector space optimization to derive the same necessary and sufficient
conditions. Specifically,
they show that the problem of finding the measurement that minimizes the
probability of a detection error can be formulated as a
generalized linear programming problem, with the positive orthant being
replaced by the positive cone of PSD matrices.
Although their approach
leads to the same conditions derived by Holevo \cite{H73}, their apparent 
suggestion that this formulation produces a standard finite-dimensional linear
programming problem is not correct, because the cone of PSD
matrices cannot be described by a finite set of linear inequalities.

In this paper, we derive the necessary and sufficient conditions for an
optimal quantum measurement in a self-contained manner, again by 
exploiting duality arguments.
 The primary
advantage of our formulation is that it readily lends itself to
efficient computational methods.
Specifically, we  show that the optimal measurement can be found by
solving a standard convex semidefinite program with $n^2$ variables,
followed by the solution of 
a set of linear equations or, at worst, a standard linear programming problem. By exploiting the many well-known
algorithms for solving semidefinite programs \cite{VB96,A91t,A92,NN94}, the optimal measurement can be
computed very efficiently in polynomial time. Furthermore, 
in contrast to the iterative
algorithm proposed by Helstrom \cite{H82} for solving the quantum
detection problem, which is only
guaranteed 
to converge to a local optimum,   algorithms based on semidefinite programming are guaranteed
to converge to the global optimum.

After a statement of the problem in  Section~\ref{sec:qd}, we derive in
Section~\ref{sec:sdp} the necessary and sufficient conditions
for the 
optimal measurement that minimizes the probability of a detection error, by
formulating our problem as a semidefinite program.
Using this formulation, in Section~\ref{sec:rone}
we prove that 
if the quantum state ensemble is a pure-state ensemble consisting of
rank-one density operators
$\rho_i=\ket{\phi_i}\bra{\phi_i}$,
then the optimal measurement is a pure-state measurement consisting of rank-one
measurement operators $\Pi_i=\ket{\mu_i}\bra{\mu_i}$.
This generalizes a previous result by Kennedy
\cite{K73}, which establishes that for {\em linearly independent}
vectors $\ket{\phi_i}$ the optimal
measurement is a (necessarily orthogonal) pure-state measurement.
We also show that for a general quantum state ensemble, the rank of
each optimal measurement operator $\Pi_i$ is no larger than the rank of
the corresponding density matrix 
$\rho_i$. In Section~\ref{sec:comp} we consider efficient iterative
algorithms that are
guaranteed to converge to the globally optimum measurement.

Throughout the paper we use the 
Dirac bra-ket notation of quantum mechanics.
In this notation, the elements of $\HH$ are ``ket" vectors,
denoted, \eg by $\ket{x} \in \HH$.  The corresponding ``bra" vector
$\bra{x}$ is the conjugate transpose of $\ket{x}$.  The inner product of two
vectors is a complex number denoted by $\braket{x}{y}$.  An outer
product of two vectors such as
$\ket{x}\bra{y}$ is a rank-one matrix, which takes
$\ket{z} \in \HH$ to $ \braket{y}{z} \ket{x}
\in \HH$.

%%%%%%%%%%%%%%%%%%%%%%%%%%%%%%%%%%%%%%%%%%%%%%%
\section{Optimal Detection of Quantum States} 
\label{sec:qd} 
%%%%%%%%%%%%%%%%%%%%%%%%%%%%%%%%%%%%%%%%%%%%%%%

Assume that a quantum channel is prepared in a 
quantum state drawn from a collection of given states.
The quantum states are represented by a set of $m$ PSD
Hermitian density operators $\{ \rho_i,1 \leq i \leq m \}$ on an $n$-dimensional
complex Hilbert 
space $\HH$. At the receiver,
a measurement is constructed, comprising $m$ 
PSD
Hermitian measurement
operators $\{\Pi_i,1 \leq i \leq m\}$ on $\HH$. The problem is to
choose the measurement operators to minimize
the probability of detection error, \ie the probability of incorrect
detection of the transmitted state.

We assume without loss of generality that the  eigenvectors of 
the density operators $\{ \rho_i,1 \leq i \leq m \}$ span\footnote{Otherwise we can transform the problem to a problem equivalent to the one considered in this paper by reformulating the problem on the subspace spanned by the eigenvectors of $\{\rho_i,1 \leq i \leq m\}$. } $\HH$. 
In this case, to constitute a measurement, the measurement  
operators $\Pi_i$  must satisfy 
\begin{equation}
\label{eq:identu}
\sum_{i=1}^m \Pi_i =I,
\end{equation}
where $I$ is the identity operator on $\HH$.

We seek the PSD measurement operators $\{\Pi_i,1 \leq i \leq m\}$ satisfying (\ref{eq:identu}) that
minimize the probability of  
detection error, or equivalently, maximize the probability of correct
detection. Given that the transmitted state is $\rho_j$, the
probability of correctly detecting the state using measurement
operators $\{\Pi_i,1 \leq i \leq m\}$ is
$\tr(\rho_j\Pi_j)$. Therefore, the probability of correct detection is
given by
\begin{equation}
\label{eq:pe}
P_d=\sum_{i=1}^mp_i\tr(\rho_i\Pi_i),
\end{equation}
where $p_i>0$ is the prior probability of $\rho_i$, with $\sum_i p_i=1$. 
Denoting by
$\B$ 
the set of Hermitian operators on $\HH$ and defining
$\rho_i'=p_i\rho_i$, our problem reduces to the maximization problem
\begin{equation}
\label{eq:max}
\max_{\Pi_i \in \B} \sum_{i=1}^m \tr(\rho_i'\Pi_i),
\end{equation}
subject to the constraints
\begin{eqnarray}
\label{eq:cond_pd}
\Pi_i & \geq & 0,\quad 1 \leq i \leq m;\\
\label{eq:cond_id}
\sum_{i=1}^m \Pi_i & = & I.
\end{eqnarray}
Denoting by $\Lambda$  the set of all ordered sets
$\Pi=\{\Pi_i\}_{i=1}^m, \Pi_i \in \B$,
satisfying (\ref{eq:cond_pd}) and (\ref{eq:cond_id}), and defining $J(\Pi)=
\sum_{i=1}^m \tr(\rho_i'\Pi_i)$,  we can express our maximization
problem as
\begin{equation}
\max_{\Pi \in \Lambda} J(\Pi).
\end{equation}
We refer to $\Lambda$ as the feasible set, and to any $\Pi \in
\Lambda$ as a feasible point.
Since $\Lambda$  is a compact set and $J(\Pi)$
is a  
continuous linear functional, there exist an optimal
$\widehat{\Pi}\in\Lambda$ and an optimal value $\widehat{J}$ defined by
\begin{equation}
\widehat{J}= J(\widehat{\Pi}) \ge J(\Pi),\quad \forall\ \Pi \in\Lambda.
\end{equation}

Equipped with the standard operations of
addition and multiplication by real numbers, 
$\B$ is an $n^2$-dimensional {\em real}
vector space.
By choosing an appropriate basis for $\B$, the problem of 
(\ref{eq:max})--(\ref{eq:cond_id}) can be put in the form of a
standard semidefinite 
programming   problem, which is a  convex
optimization problem; for a detailed treatment of semidefinite programming problems
see, \eg  \cite{A91t,A92,NN94,VB96}. Rather than
relying on results that 
are scattered throughout the literature in various forms,
in what follows we present a 
self-contained and direct
derivation of the necessary and sufficient conditions
for the optimal measurement. As we will see, this derivation also leads to efficient
methods for computing the optimal measurement in cases in which an  analytical
solution is not known. 

In the next section we derive the necessary and sufficient conditions on the
measurement operators by formulating a
{\em dual problem}. The dual problem will also be used in
Section~\ref{sec:comp} to develop efficient computational algorithms.

%%%%%%%%%%%%%%%%%%%%%%%%%%%%%%%%%%%%%%%%%%%%%%%
\section{Dual Problem Formulation} 
\label{sec:sdp} 
%%%%%%%%%%%%%%%%%%%%%%%%%%%%%%%%%%%%%%%%%%%%%%%

Our objective is to formulate a {\em dual problem} whose optimal value
serves as a 
certificate for $\widehat{J}$. Specifically, we will
formulate a minimization problem of the form $\min_X T(X)$ for some 
linear functional $T$ such that for all feasible
values of 
$X \in \B$, \ie values of $X \in \B$ that satisfy a certain set of
constraints, and 
for any $\Pi \in \Lambda$, we shall have $T(X) \geq J(\Pi)$.
The
dual problem therefore provides an upper bound on the optimal value of the
original (primal) problem.
In addition we would like the minimal value of $T$, denoted $\widehat{T}$, to
be equal to  $\widehat{J}$. 
The equality $\widehat{J}=\widehat{T}$ will then lead to
conditions of optimality on the measurement operators. Furthermore, in
this case, instead of solving the primal
problem, we can find $\widehat{J}$ and the optimal measurement by solving the dual
problem, which turns out to have far fewer decision variables.

%%%%%%%%%%%%%%%%%%%%%%%%%%%%%%%%%%%%%%%%%%%%%%%
\subsection{Constructing The Dual Problem} 
\label{sec:dp} 
%%%%%%%%%%%%%%%%%%%%%%%%%%%%%%%%%%%%%%%%%%%%%%%

A general method for deriving a dual problem is to invoke
the separating hyperplane theorem \cite{L68}, which states that two disjoint convex 
sets\footnote{A set $C$ is convex if for any $x,y \in C$, $\alpha
x+(1-\alpha)y \in C$ for all $\alpha \in [0,1]$.} can always be separated by a hyperplane. We will take one convex set to be the  point $0$, and then carefully construct another convex set that does not 
contain $0$. This set will capture the 
 equality constraints in the primal problem and the fact
that for any primal feasible point, the value of the primal function
is no larger than the optimal value.
The dual variables will then emerge from the parameters of the separating
hyperplane. 

In our problem we have one equality constraint,
$\sum_{i=1}^m\Pi_i=I$, and we know that $\widehat{J} \geq J(\Pi)$. Our constructed convex  set will accordingly consist of matrices 
of the form $-I+\sum_{i=1}^m\Pi_i$ where $\Pi_i \in \B$ and $\Pi_i
\geq 0$, and scalars of the form
$r-J(\Pi)$ where $r>\widehat{J}$.
We thus consider the 
$(n^2+1)$-dimensional real
vector space 
\[\LL=\B\times\R=\{(S,x):\ \ S\in \B,\ x\in\R\},\]
where $\R$ denotes the reals, with inner product defined by
\begin{equation}
\label{eq:ip}
\inner{(W,y)}{(S,x)}=\tr(WS)+yx.
\end{equation}
Note that since $W,S \in \B$, $\tr(WS) \in \R$. 

We now define the subset $\Omega$  of $\LL$ by
\begin{equation}
 \Omega=\left\{\left(-I+\sum_{i=1}^m\Pi_i,\;
r-\sum_{i=1}^m\tr(\Pi_i\rho'_i)\right):\ \ 
\Pi_i\in \B,\ \Pi_i\ge0,\ r\in\R,\ r>\widehat{J}\right\}.
\end{equation}
It is easily verified that $\Omega$ is convex, and $0\not\in\Omega$. 
Therefore, by the
separating hyperplane theorem, there exists
a {\em nonzero} vector $(Z,a) \in \LL$ such that 
$\inner{(Z,a)}{(Q,b)} \geq 0$
for all $(Q,b)\in\Omega$, \ie 
\begin{equation}
\label{eq:hyperplane}
\tr\left(Z\left(-I+\sum_{i=1}^m\Pi_i\right)\right)+
a\left(r-\sum_{i=1}^m\tr(\Pi_i\rho'_i)\right)\ge0
\end{equation}
for all $\Pi_i\in \B$ and $r\in\R$ such that $\Pi_i\ge0$, $r>\widehat{J}$.
It will turn out that the hyperplane parameters $(Z,a)$ define the optimal dual
point. We first show that these parameters have to satisfy certain
constraints, which lead to the formulation of the dual problem. 

Note that
(\ref{eq:hyperplane}) with $\Pi_i=0$, $r\to \widehat{J}$ implies 
\begin{equation}
\label{eq:a}
a\widehat{J}\ge\tr(Z).
\end{equation}
Similarly, (\ref{eq:hyperplane}) with $r=\widehat{J}+1$, $\Pi_j=0$ for
$j\neq i$, 
$\Pi_i=txx'$ where $x\in\R^n$ is fixed and $t\to+\infty$ yields
$x'(Z-a\rho'_i)x\ge0$. Since $x$ and $i$ are arbitrary, this implies
\begin{equation}
\label{eq:Z}
Z\ge a\rho'_i,\quad 1 \leq i \leq m.
\end{equation}
With $\Pi_i=0$, $r\to+\infty$, (\ref{eq:hyperplane}) implies
$a\ge0$. If  $a=0$,
then (\ref{eq:Z}) yields $Z\ge0$, and
(\ref{eq:a}) yields $0\ge\tr(Z)$, which together means $Z=0$.
However, this would contradict the assumption that $(Z,a) \neq 0$.
Therefore we conclude that $a>0$, and define
$\hx=Z/a$. Then (\ref{eq:a}) implies that 
\begin{equation}
\label{eq:tx}
T(\hx)\le \widehat{J},
\end{equation}
where $T(X)=\tr(X)$, and 
(\ref{eq:Z})
implies  that $\hx \geq \rho_i'$ for $1 \leq i \leq m$.
Let $\Gamma$ be the
set of $X \in \B$ satisfying $X \geq \rho'_i,1 \leq i \leq m$.
Then for any
 $X\in\Gamma$, 
$\Pi \in\Lambda$, we have
\begin{equation}
\label{eq:wduality}
T(X)-J(\Pi)=\sum_{i=1}^m\tr \bl \Pi_i(X-\rho'_i) \br\ge 0.
\end{equation}
Since $\hx \in \Gamma$, from (\ref{eq:tx}) and 
(\ref{eq:wduality}) we conclude that 
$T(\hx)=\widehat{J}$.

Thus we have proven that the dual problem associated 
with (\ref{eq:max})--(\ref{eq:cond_id}) is 
\begin{equation}
\label{eq:min}
\min_{X \in \B} T(X),
\end{equation}
where $T(X)=\tr(X)$,
subject to
\begin{equation}
\label{eq:condx}
X \geq \rho'_i, \quad 1 \leq i \leq m.
\end{equation}
Furthermore, we have shown that 
there exists  an optimal
$\hx \in\Gamma$ and an optimal 
value $\widehat{T}$ defined by
\begin{equation}
\widehat{T}=  T(\widehat{X})\le T(X),\quad \forall\ X\in\Gamma,
\end{equation}
such that
\begin{equation}
\label{eq:optimal}
\widehat{T}=\widehat{J}. 
\end{equation}

%%%%%%%%%%%%%%%%%%%%%%%%%%%%%%%%%%%%%%%%%%%%%%%
\subsection{Optimality Conditions} 
\label{sec:opt} 
%%%%%%%%%%%%%%%%%%%%%%%%%%%%%%%%%%%%%%%%%%%%%%%

Let $\hpi_i$ denote the optimal measurement operators that
maximize (\ref{eq:max}) 
subject to (\ref{eq:cond_pd}) and (\ref{eq:cond_id}), and let 
$\hx$ denote the optimal $X$ that minimizes (\ref{eq:min})
subject to (\ref{eq:condx}). Then from
(\ref{eq:optimal}) it follows  that
\begin{equation}
\label{eq:duality}
\sum_{i=1}^m\tr \bl \hpi_i(\hx-\rho'_i)\br= 0.
\end{equation}
Since $\hx \geq \rho_i'$ and $\Pi_i \geq 0$, 
(\ref{eq:duality}) is satisfied if and only if 
\begin{equation}
\label{eq:op_cond}
(\hx-\rho_i')\hpi_i = \hpi_i(\hx-\rho_i')=0, \quad 1 \leq i \leq m.
\end{equation}

Once we find the optimal $\hx$ that minimizes
the dual problem (\ref{eq:min}), the constraint (\ref{eq:op_cond})
is a necessary and sufficient 
condition on the optimal measurement operators $\hpi_i$. We have already
seen that this condition is necessary. To show that it is sufficient,
we note that if a set of measurement operators $\Pi_i$ satisfies
(\ref{eq:op_cond}), then 
$\sum_{i=1}^m\tr \bl \Pi_i(\hx-\rho'_i)\br= 0$ so that
$J(\Pi)=T(\hx)=\widehat{J}$.

Since the dual problem involves many fewer decision
variables than the primal maximization problem,
it is advantageous to 
solve the dual problem and then use 
 (\ref{eq:op_cond}) to determine the optimal measurement operators,
rather than solving the primal problem directly.
In Section~\ref{sec:comp} we develop efficient algorithms that follow
this strategy.

Using (\ref{eq:identu}), 
(\ref{eq:op_cond}) and (\ref{eq:condx}) leads to the conditions
\begin{eqnarray}
\label{eq:op_cond1}
\sum_{i=1}^m \rho_i'\hpi_i & = & \sum_{i=1}^m \hpi_i\rho_i'; \\
\label{eq:op_cond2}
\sum_{i=1}^m \rho_i'\hpi_i & \geq & \rho_j', \quad 1 \leq i \leq m.
\end{eqnarray}
Thus, any optimal measurement $\hpi=\{\hpi_i\}_{i=1}^m$ must satisfy (\ref{eq:op_cond1}) and
(\ref{eq:op_cond2}). These conditions are also
derived in 
\cite{YKL75,H73}. However, as noted in the Introduction,
the approach taken here lends itself
to fast iterative algorithms, as we will see in
Section~\ref{sec:comp}, and also provides additional insight into
the optimal measurement operators, as we show in
Section~\ref{sec:rone}. 

In \cite{YKL75} it was established that the conditions (\ref{eq:op_cond1}) and
(\ref{eq:op_cond2}) together with (\ref{eq:cond_pd})
and (\ref{eq:cond_id})  are  also sufficient. For completeness, we
repeat the argument here.
Suppose that the measurement operators $\hpi_i$ satisfy
(\ref{eq:op_cond1}) and 
(\ref{eq:op_cond2}). Then $\hx=\sum_{i=1}^m\hpi_i \rho_i' \in \Gamma$.
It then follows from (\ref{eq:wduality}) that for any
set of measurement operators $\Pi_i \in \Lambda$,
\begin{equation}
\sum_{i=1}^m\tr \bl \Pi_i\rho'_i\br \leq
\tr(\hx)=\sum_{i=1}^m\tr(\hpi_i\rho_i')
\end{equation}
with equality for $\Pi_i=\hpi_i$. Therefore the measurement operators
$\hpi_i$ are optimal.

We summarize our results in the following theorem:
\begin{theorem}
\label{thm:dual}
Let $\{\rho_i,1 \leq i \leq m\}$ denote a set of density operators with
prior probabilities $\{p_i>0,1 \leq i \leq m\}$, and let
$\{\rho'_i=p_i\rho_i,1 \leq i \leq m\}$. 
Let  $\Lambda$ denote the set  of all ordered sets of Hermitian measurement
operators $\Pi=\{\Pi_i\}_{i=1}^m$ that satisfy $\Pi_i \geq 0$ and
$\sum_{i=1}^m \Pi_i=I$, and let  $\Gamma$ denote the set of
Hermitian matrices $X$ such 
that $X \geq \rho_i,1 \leq i \leq m$.
Consider the problem $\max_{\Pi \in \Lambda} J(\Pi)$ and the dual problem
$\min_{X \in \Gamma} T(X)$, where $J(\Pi)=\sum_{i=1}^m
\tr(\rho_i'\Pi_i)$ and 
$T(X)=\tr(X)$.
Then
\begin{enumerate}
\item For any $X \in \Gamma$ and $\Pi \in \Lambda$, $T(X) \geq J(\Pi)$.
\item There is an optimal $\Pi$, denoted $\hpi$, such that
$\widehat{J}=J(\hpi) \geq J(\Pi)$ for any $\Pi \in \Lambda$;
\item There is an optimal $X$, denoted $\hx$, such that
$\widehat{T}=T(\hx) \leq T(X)$ for any $X \in \Gamma$;
\item $\widehat{T}=\widehat{J}$;
\item Given $\hx$, a necessary and sufficient condition on the
optimal measurement operators $\hpi_i$ is
$(\hx-\rho_i')\hpi_i=0,1 \leq i \leq m$.
\end{enumerate}
\end{theorem}

%%%%%%%%%%%%%%%%%%%%%%%%%%%%%%%%%%%%%%%%%%%%%%%
\section{Rank-One Ensembles} 
\label{sec:rone} 
%%%%%%%%%%%%%%%%%%%%%%%%%%%%%%%%%%%%%%%%%%%%%%%

Suppose now that the density operators $\rho_i$
are rank-one operators
 of the
form $\rho_i=\ket{\phi_i}\bra{\phi_i}$ for some $\ket{\phi_i} \in \HH$.
In this case, it seems intuitively plausible that the optimal measurement will
consist of rank-one measurement operators 
of the form
$\hpi_i=\ket{\mu_i}\bra{\mu_i}$ for some $\ket{\mu_i} \in \HH$.

There are some particular cases in which an analytical
solution to the quantum
detection problem is known \cite{H76,CBH89,OBH96,BKMO97,EF01}.
In all of these cases, when the density operators are rank-one
operators, the  optimal measurement  also  has rank one.
In the special case in which the vectors $\ket{\phi_i}$ are
{\em linearly independent},  Kennedy
\cite{K73} showed that the optimal measurement is always a rank-one
measurement. 
However, this implication has not been proven in the general
case.
Using the conditions for optimality we derived in the
previous section, we now prove this implication for an arbitrary
rank-one ensemble. 

We have seen that the optimal measurement operators $\hpi_i$ can be 
determined by solving (\ref{eq:op_cond}), where $\hx$ is the optimal
matrix that minimizes (\ref{eq:min}) subject to (\ref{eq:condx}).
Thus the measurement operators $\hpi_i$ must lie in the null space  of
$\hx-\rho_i'$, denoted 
$\N(\hx-\rho_i')$, and consequently $\mbox{rank}(\hpi_i) \leq \dim\bl \N(\hx-\rho_i') \br$.  

Since  $\hx \geq \rho_i,1 \leq i \leq m$, it follows that $\hx$
is positive 
definite on $\HH$.
Indeed, since the eigenvectors of the matrices $\rho_i$ span $\HH$, for any $h \in \HH$ there exists an $i$ such that
$\braket{h}{\rho'_i|h}>0$, which implies that $\braket{h}{\hx|h}>0$ for
any $h \in \HH$, so that $\N(\hx)=\{0\}$.
Now, for any two matrices $Z_1$ and $Z_2$, $\mbox{rank}(Z_1+Z_2) \geq
\mbox{rank}(Z_1)-\mbox{rank}(Z_2)$, so that
\begin{equation}
\label{eq:dim2}
\dim(\N(Z_1+Z_2)) \leq \dim(\N(Z_1))+\mbox{rank}(Z_2).
\end{equation}
With $Z_1=\hx$ and
$Z_2=-\rho_i'$, (\ref{eq:dim2}) yields
\begin{equation}
\label{eq:dim3}
\dim\bl \N(\hx-\rho_i') \br
\leq \mbox{rank}(\rho_i')=\mbox{rank}(\rho_i),
\end{equation}
and 
\begin{equation}
\label{eq:rank}
\mbox{rank}(\hpi_i) \leq \dim\bl \N(\hx-\rho_i')  \br \leq 
\mbox{rank}(\rho_i),\quad 1 \leq i \leq m.
\end{equation}

In the special case in which the operators
$\rho_i=\ket{\phi_i}\bra{\phi_i}$ have rank-one, it 
follows immediately from (\ref{eq:rank}) that the optimal measurement
operators  also have rank-one,
so that they have the form
$\hpi_i=\ket{\mu_i}\bra{\mu_i}$ for some $\ket{\mu_i} \in \HH$.

If in addition the vectors $\{\ket{\phi_i},1 \leq i \leq m\}$ are
linearly independent, 
then the vectors $\{\ket{\mu_i},1 \leq i \leq m\}$ must also be
linearly independent since 
$\sum_{i=1}^m \ket{\mu_i}\bra{\mu_i}$ is equal to the identity on $\HH$, where now $\HH$ is the 
$m$-dimensional space spanned by the vectors
$\ket{\phi_i}$. Then, for $1 \leq j \leq m$,
\begin{equation}
\ket{\mu_j}=
\sum_{i=1}^m \braket{\mu_i}{\mu_j}\ket{\mu_i}.
\end{equation}
Since the vectors $\ket{\mu_i}$  are linearly independent, we must have
that $\braket{\mu_i}{\mu_j}=\delta_{ij}$  so that the vectors
$\ket{\mu_i}$  are mutually orthonormal.
We therefore recover the statement by Kennedy \cite{K73}, that 
for a pure-state  ensemble with linearly independent vectors,
the optimal measurement is an
orthogonal pure-state measurement.

We summarize our results in the following theorem:
\begin{theorem}
\label{thm:rank1}
Let $\{\rho_i,1 \leq i \leq m\}$ be a quantum state ensemble
consisting of density operators $\rho_i$ with prior probabilities $p_i>0$.
Then the optimal measurement 
consists of measurement operators $\{\Pi_i,1 \leq i \leq m\}$ with 
$\mbox{rank}(\Pi_i) \leq \mbox{rank}(\rho_i)$. In particular 
if  $\{\rho_i=\ket{\phi_i}\bra{\phi_i},1 \leq i \leq m\}$ is a
pure-state quantum  
ensemble, then the optimal measurement is a
pure-state measurement consisting of measurement operators of the form 
$\{\Pi_i=\ket{\mu_i}\bra{\mu_i},1 \leq i \leq m\}$.
\end{theorem}

%%%%%%%%%%%%%%%%%%%%%%%%%%%%%%%%%%%%%%%%%%%%%%%
\section{Computational Aspects} 
\label{sec:comp} 
%%%%%%%%%%%%%%%%%%%%%%%%%%%%%%%%%%%%%%%%%%%%%%%

In the general case there is no closed-form analytical solution to 
the maximization problem (\ref{eq:max}) or the minimization problem
(\ref{eq:min}). However, since 
(\ref{eq:min}) is a convex optimization problem, there are very
efficient methods for solving (\ref{eq:min}). In particular, the optimal matrix $\hx$ minimizing
$\tr(X)$ subject to (\ref{eq:condx}) can be computed in Matlab using
the linear matrix inequality (LMI) Toolbox. A 
convenient interface for using the LMI toolbox is the 
Matlab package\footnote{This software was created by A.~Megretski,
C-Y.~Kao, U.~J\"{o}nsson and A.~Rantzer and is available at
\texttt{http://www.mit.edu/cykao/home.html}.}  IQC$\beta$.

 Once we  determine $\hx$, 
the optimal measurement operators
$\hpi_i$ can be computed using  (\ref{eq:op_cond}), (\ref{eq:cond_pd})
and (\ref{eq:cond_id}).  
Specifically, from (\ref{eq:op_cond}) and (\ref{eq:cond_pd}) it
follows that $\hpi_i$ can be expressed as 
\begin{equation}
\label{eq:hipa}
\hpi_i=\sum_{j=1}^t a_{ij} \ket{q_{ij}}\bra{q_{ij}}, 
\end{equation}
where $a_{ij} \geq 0$, $t \leq \mbox{rank}(\rho_i)$, and 
the vectors $\ket{q_{ij}},1 \leq j \leq t$ are
orthonormal and span
$\N(\hx-\rho_i')$. To determine the vectors $\ket{q_{ij}}$ we may use
the   
eigendecomposition of $\hx-\rho'_i$.

To satisfy
(\ref{eq:cond_id}) we must have 
\begin{equation}
\label{eq:sumt}
\sum_{i=1}^m \sum_{j=1}^t a_{ij} \ket{q_{ij}}\bra{q_{ij}}=I.
\end{equation}
Let $\ket{e}=\mvec(I)$ and $\ket{y_{ij}}=\mvec(\ket{q_{ij}}\bra{q_{ij}})$,
where $\ket{v}=\mvec(V)$ denotes the vector obtained by stacking the columns of
$V$. Then we can express (\ref{eq:sumt}) as 
\begin{equation}
\label{eq:veca}
Y\ket{a}=\ket{e},
\end{equation}
where $Y$ is the matrix of columns $\ket{y_{ij}}$ and $\ket{a}$ is the
vector with components $a_{ij}$.
If the matrix $Y$ has full column rank, then the unique solution to
(\ref{eq:veca}) is 
\begin{equation}
\label{eq:afr}
\ket{a}=(Y^*Y)^{-1}Y^*\ket{e}.
\end{equation}
In the general case, $Y$ will not have full column rank and  there will
be many solutions $\ket{a}$ to (\ref{eq:veca}). 
Each such vector defines a corresponding set of optimal measurement
operators $\hpi_i$ via
(\ref{eq:hipa}).
To find a unique solution we may seek the vector\footnote{The inequality is to be understood as a component-wise
inequality.} $\ket{a} \geq 0$ that
satisfies (\ref{eq:veca}), and such that 
$\sum_{i=1}^m\tr(\hpi_i)=\sum_{i=1}^m \sum_{j=1}^t a_{ij}$ is minimized. 
Our problem therefore reduces to 
\begin{equation}
\label{eq:mint}
\min \braket{1}{a}
\end{equation}
where $\ket{1}$ denotes the vector with  components that are all equal
$1$, 
subject to
\begin{eqnarray}
\label{eq:constlp}
Y\ket{a}=\ket{e}; \nonumber  \\
\ket{a} \geq 0.
\end{eqnarray}
The problem of (\ref{eq:mint})--(\ref{eq:constlp}) is just a standard
linear programming  problem that can be solved very efficiently using
standard linear programming tools \cite{BT97}, for example the LMI toolbox in
Matlab.
%\footnote{Note that the Matlab LMI solver solves the dual linear
%probelm $\min -\inner{e}{y}$ sbject to $Y^*\ket{y} \leq
%\ket{1}$. Theo ptiaml value of $\ket{a}$ and $\ket{y}$ are related
%through $\inner{\ket{a}}{Y^*\ket{y}-\ket{1}}=0$}

%%%%%%%%%%%%%%%%%%%%%%%%%%%%%%%%%%%%%%%%%%%%%%%
\subsection{Example} 
\label{sec:example} 
%%%%%%%%%%%%%%%%%%%%%%%%%%%%%%%%%%%%%%%%%%%%%%%

We now consider an example illustrating the computational steps
involved in
computing the optimal measurement.

Consider the case in which the ensemble consists of 
$3$ rank-one density operators
$\rho_i=\ket{\phi_i}\bra{\phi_i},1 \leq i \leq 3$  where
\begin{equation}
\label{eq:ex}
\ket{\phi_{1}}=\left[
\begin{array}{r}
1  \\
0
\end{array}
\right],\,\,\,
\ket{\phi_{2}}=\frac{1}{\sqrt{2}}\left[
\begin{array}{r}
1  \\
1
\end{array}
\right],\,\,\,
\ket{\phi_{3}}=\left[
\begin{array}{r}
0  \\
1
\end{array}
\right],\,\,\,
\end{equation}
with prior probabilities
\begin{equation}
\label{eq:pex}
p_1=0.1,\,\,\,p_2=0.6,\,\,\,p_3=0.3.
\end{equation}
To find the optimal measurement operators, we first find the optimal
matrix $\hx$ that minimizes $\tr(X)$ subject to $X \geq
\rho'_i$ with $\rho'_i=p_i\rho_i$.
The matrix $\hx$ is computed using the IQC toolbox on Matlab.
To this end we generate the following code:
\begin{singlespace}
\begin{tabular}{ll}
\texttt{>> abst\_init\_lmi} & \% Initializing
the LMI toolbox\\ 
\texttt{>> X=symmetric(2);} & \% Defining a symmetric $2 \times 2$
variable $X$ \\
\texttt{>> } $\mathtt{ X>p1*R1;}$ & \% 
Imposing the inequality constraints: \\
\texttt{>> } $\mathtt{ X>p2*R2;}$ &  \% Here $\mathtt{p1}=p_1,
\mathtt{p2}=p_2, \mathtt{p3}=p_3$ and \\
\texttt{>> } $\mathtt{ X>p3*R3;}$ & \% $\mathtt{R1}=\rho_1,
\mathtt{R2}=\rho_2, \mathtt{R3}=\rho_3$\\
\texttt{>> lmi\_mincx\_tbx(trace(X));} \hspace*{0.4in} & \% Minimizing $\tr(X)$
subject to the constraints\\
\texttt{>> X=value(X)} & \% Getting the optimal value of $X$ 
\end{tabular}
\end{singlespace}
\vspace*{0.1in}
\noindent The optimal $\hx$ is given by 
\begin{equation}
\label{eq:hx}
\hx=\left[
\begin{array}{ll}
0.352 & 0.217  \\
0.217 & 0.434 
\end{array}
\right].
\end{equation}
Using the eigendecomposition of $\hx-\rho'_i$ we conclude that, as we
expect from Theorem~\ref{thm:rank1}, 
$\N(\hx-\rho'_i)$  has dimension $1$ for each $i$ and is
spanned by the vector $\ket{q_i}$ where 
\begin{equation}
\ket{q_{1}}=\left[
\begin{array}{r}
-0.833  \\
0.554
\end{array}
\right],\,\,\,
\ket{q_{2}}=\frac{1}{\sqrt{2}}\left[
\begin{array}{r}
0.850  \\
0.527
\end{array}
\right],\,\,\,
\ket{q_{3}}=\left[
\begin{array}{r}
-0.525  \\
0.851
\end{array}
\right].\,\,\,
\end{equation}
The optimal measurement operators are therefore given by
$\hpi_i=a_i\ket{q_i}\bra{{q_i}}=\ket{\mu_i}\bra{\mu_i}$ with
$\ket{\mu_i}=\sqrt{a_i}\ket{q_i}$ and $a_i$ denoting the $i$th component of
$\ket{a}$. From (\ref{eq:veca}) $\ket{a}$ must satisfy
\begin{equation}
\label{eq:ao}
\left[
\begin{array}{rrr}
0.693 & 0.722   &0.276 \\
-0.461 & 0.448  &  -0.447\\
-0.461 & 0.448  & -0.447\\
0.306 & 0.278 &  0.724
\end{array}
\right]
\left[
\begin{array}{c}
a_1 \\
a_2 \\
a_3
\end{array}
\right]=
\left[
\begin{array}{c}
1 \\
0 \\
0 \\
1
\end{array}
\right].
\end{equation}
Since the matrix in (\ref{eq:ao}) has full column rank, there is a unique
solution 
\begin{equation}
\ket{a}=\left[
\begin{array}{c}
0.007 \\
0.999 \\
0.994
\end{array}
\right].
\end{equation}
The optimal measurement vectors are then given by
$\ket{\mu_i}=\sqrt{a_i}\ket{q_i}$ which yields
\begin{equation}
\label{eq:mex}
\ket{\mu_1}=\left[
\begin{array}{r}
-0.067  \\
0.046
\end{array}
\right],\,\,\,
\ket{\mu_{2}}=\left[
\begin{array}{r}
0.849  \\
0.527
\end{array}
\right],\,\,\,
\ket{\mu_{3}}=\left[
\begin{array}{r}
-0.524  \\
0.849
\end{array}
\right].
\end{equation}

We can immediately verify that the measurement operators
$\hpi_i=\ket{\mu_i}\bra{\mu_i}$ with $\ket{\mu_i}$ given by
(\ref{eq:mex}) together with $\hx$ given by (\ref{eq:hx}) satisfy the necessary
and sufficient conditions 
 (\ref{eq:cond_pd}), (\ref{eq:cond_id}) and 
(\ref{eq:op_cond}). Furthermore, we have that the probability of
correct detection is given by
\begin{equation}
\label{eq:pdo}
\tr(\hx)=\sum_{i=1}^m p_i\tr(\hpi \rho_i)=0.78.
\end{equation}

In Fig.~\ref{fig:ex} we plot the weighted state vectors $\ket{\psi_i}=
\sqrt{p_i}\ket{\phi_i}$ given by
(\ref{eq:ex}) and (\ref{eq:pex}), together with
 the optimal measurement vectors
$\ket{\mu_i}$ given by (\ref{eq:mex}). For comparison, we also plot
the least-squares measurement vectors $\ket{\chi_i}$ which are given by
\cite{EF01} 
\begin{equation}
\label{eq:lsm}
\ket{\chi_i}=(\Psi\Psi^*)^{-1/2}\ket{\psi_i},
\end{equation}
where $\Psi$ is the matrix of columns $\ket{\psi_i}$
and
$(\cdot)^{1/2}$ is the unique symmetric square root of the
corresponding matrix. Note, that since the vectors
$\ket{\phi_i}$ span $\HH$, $\Psi\Psi^*$ is invertible.
The probability of correct detection using the least-squares
measurement vectors is 
$\sum_{i=1}^m p_i|\braket{\chi_i}{\phi_i}|^2=0.71$. As we expect, this
probability is smaller than the probability of correct detection using
the optimal measurement vectors which from (\ref{eq:pdo}) is equal to $0.78$.
\setlength{\unitlength}{.25in}
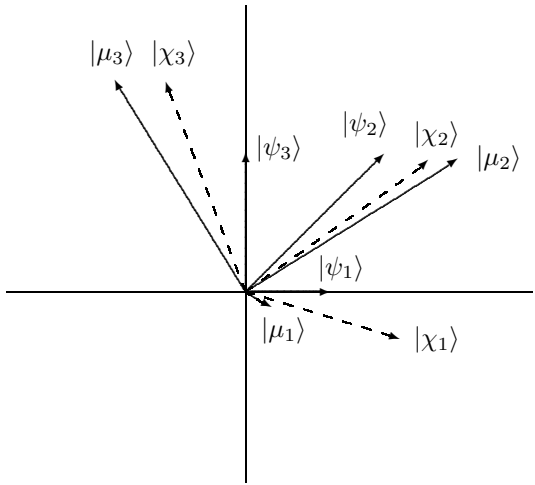
\begin{figure}[h]
\begin{center}
\begin{picture}(15,11)(0,0)
%\put(0,0){\framebox(15,11){}}

\put(0,-1.5){
\put(7,6){\line(0,1){6}}
\put(7,6){\line(1,0){6}}
\put(7,6){\line(0,-1){4}}
\put(7,6){\line(-1,0){5}}

% Original

\put(7,6){\curve(0,0,1.58,0)}
\put(8.58,6){\vector(1,0){0.2}}
\put(7,6){\curve(0,0,2.74,2.74)}
\put(9.7,8.7){\vector(1,1){0.2}}
\put(7,6){\curve(0,0,0,2.74)}
\put(7,8.74){\vector(0,1){0.2}}

% Optimal

\put(7,6){\curve(0,0,0.35,-0.23)}
\put(7.35,5.77){\vector(2,-1){0.2}}
\put(7,6){\curve(0,0,4.25,2.64)}
\put(11.25,8.66){\vector(3,2){0.2}}
\put(7,6){\curve(0,0,-2.62,4.24)}
\put(4.46,10.05){\vector(-1,2){0.2}}

% LS

\put(7,6){\dashline{0.2}(0,0)(0.35,-0.23)}

\put(7,6){\dashline{0.2}(0,0)(3.03,-0.92)}
\put(10.03,5.08){\vector(3,-1){0.2}}
\put(7,6){\dashline{0.2}(0,0)(3.65,2.59)}
\put(10.62,8.56){\vector(1,1){0.2}}
\put(7,6){\dashline{0.2}(0,0)(-1.59,4.18)}
\put(5.43,10.1){\vector(-1,3){0.1}}

%\put(7,6){\curve(0,0,3.03,-0.92)}
%\put(10.03,5.08){\vector(3,-1){0.2}}
%\put(7,6){\curve(0,0,3.65,2.59)}
%\put(10.62,8.56){\vector(1,1){0.2}}
%\put(7,6){\curve(0,0,-1.59,4.18)}
%\put(5.43,10.1){\vector(-1,3){0.1}}

\put(9,6.5){\makebox(0,0){\small $\ket{\psi_1}$}}
\put(7.8,5.2){\makebox(0,0){\small $\ket{\mu_1}$}}
\put(11,5){\makebox(0,0){\small $\ket{\chi_1}$}}
\put(7.7,9){\makebox(0,0){\small $\ket{\psi_3}$}}
\put(4.2,11){\makebox(0,0){\small $\ket{\mu_3}$}}
\put(5.5,11){\makebox(0,0){\small $\ket{\chi_3}$}}
\put(9.5,9.5){\makebox(0,0){\small $\ket{\psi_2}$}}
\put(11,9.3){\makebox(0,0){\small $\ket{\chi_2}$}}
\put(12.3,8.8){\makebox(0,0){\small $\ket{\mu_2}$}}

}
\end{picture}
\caption{Illustration of the optimal measurement vectors. The weighted
state vectors are $\ket{\psi_i}=\sqrt{p_i}\ket{\phi_i}$ where the
vectors $\ket{\phi_i}$ and the probabilities $p_i$ are given by
(\ref{eq:ex}) and (\ref{eq:pex}), respectively.  The optimal
measurement vectors $\ket{\mu_i}$ are given by (\ref{eq:mex}). The
least-squares measurement vectors $\ket{\chi_i}$ are plotted in dashed
lines for comparison, and are given by (\ref{eq:lsm}).}
\label{fig:ex}
\end{center}
\end{figure}

\newpage

\begin{singlespace}
%\bibliography{paper}
%\bibliographystyle{IEEEbib.bst}

\end{singlespace}
\end{document}